\documentstyle[11pt]{article}

\makeatletter
\@addtoreset{equation}{section}
\makeatother

\begin{document}

\def \beq{\begin{equation}}
\def\eeq{\end{equation}}
\newtheorem{theo}{Theorem}
\newtheorem{defi}{Definition}
\newtheorem{prop}{Proposition}

\begin{titlepage}

\title{On discrete 3-dimensional equations associated with
the local Yang-Baxter relation}

\author{R.M. Kashaev\thanks{On leave of absence from
St. Petersburg Branch of the Steklov Mathematical Institute,
Fontanka 27, St. Petersburg 191011, RUSSIA}\\ \\
Laboratoire de Physique Th\'eorique
{\sc enslapp}\thanks{URA 14-36 du CNRS,
associ\'ee \`a l'E.N.S. de Lyon, au LAPP d'Annecy
et \`a l'Universit\`e de Savoie}
\\
ENSLyon,
46 All\'ee d'Italie,\\
69007 Lyon, FRANCE\\
E-mail: {\sf rkashaev\@@enslapp.ens-lyon.fr}}

\date{December 1995}

\maketitle

\abstract{The local Yang-Baxter equation (YBE), introduced by Maillet
and Nijhoff, is a proper generalization to 3 dimensions of the zero
curvature relation. Recently, Korepanov has constructed an infinite
set of integrable 3-dimensional lattice models, and has related them
to solutions to the local YBE. The simplest Korepanov's model is related
to the star-triangle
relation in the Ising model. In this paper the corresponding discrete
equation is derived. In the continuous limit it leads to a differential
3d equation, which is symmetric with respect to all permutations of the
three coordinates. A similar analysis of the star-triangle transformation
in electric
networks leads to the discrete bilinear equation of Miwa, associated with
the BKP hierarchy.

 Some related operator solutions to the tetrahedron equation are also
constructed.}
\vskip 2cm

\rightline{{\small E}N{\large S}{\Large L}{\large A}P{\small P}-L-569/95}

\end{titlepage}

\section{Introduction}

The Yang-Baxter equation (YBE) \cite{Y,B1} plays the central role in the
formalism of the quantum inverse scattering method of solving integrable
2d quantum systems \cite{F}. The corresponding dynamical equations can be
represented as
a zero-curvature condition (Lax equation) for some auxiliary linear problem.

 The notion of the local YBE has been introduced by Maillet and Nijhoff in
\cite{MN1,MN2} as a proper 3-dimensional generalization of the zero-curvature
condition.
Korepanov in \cite{K1,K2} has pushed forward this idea by constructing an
infinite set of discrete
3-dimensional integrable systems, and relating them
to solutions to the local YBE, the simplest solution being equivalent to
the star-triangle relation (STR) in the 2-dimensional Ising model.

In this letter first we consider the star-triangle transformation in
electric networks, and show that, being
interpreted as the local YBE, it leads to the integrable discrete bilinear
equation,
first discovered by Miwa in \cite{M}. This is the content of Sect.~\ref{sec1}.
 Then, in Sect.~\ref{secising}, in the same way we analyse the Korepanov's
model associated with the STR in the Ising model. The corresponding
integrable discrete
equation in the continuous limit
leads to the 3-dimensional non-linear differential equation, the symmetry
group of which contains all permutations of three coordinates.
In Sect.~\ref{secte}
some operator solutions to the tetrahedron equation, related with the
abovementioned models, are constructed.

\section{The star-triangle transformation in electric networks and
Miwa's equation}\label{sec1}

The star-triangle transformation in electric networks is historically
the first transformation of this kind known since the last century.
It is given by the following formulae:
\beq\label{stt}
r_jr'_j=r'_1r'_2+r'_2r'_3+r'_3r'_1=r_1r_2r_3/(r_1+r_2+r_3),\quad j=1,2,3,
\eeq
where the resistances $\{r'_j\}$ and $\{r_j\}$ correspond to the star
and triangle, respectively. There is a STR in the usual sense \cite{B},
underlying
this transformation. Define a Gaussian exponent:
\beq\label{wrq}
w(r|q)=\exp(-iq^2/2r),\quad q,r\in R.
\eeq
Then, it is easy to check that the following STR holds:
\begin{eqnarray}\label{str1}
\int_{-\infty}^{+\infty}dq''w(r'_3|q-q'')w(r'_2|q'')w(r'_1|q''-q')\nonumber\\
=\sqrt{2\pi r'_1r'_3/ir_2}w(r_1|q)w(r_2|q-q')w(r_3|q'),
\end{eqnarray}
where the parameters $\{r'_j\}$, $\{r_j\}$ satisfy
relations (\ref{stt}). The STR can be rewritten in the operator form as
follows.

Let $C[\hat p,\hat q]$ be the Heisenberg algebra generated by Hermitian
operators $\hat p$ and $\hat q$, satisfying the Heisenberg permutation
relation:
\beq
\hat q\hat p-\hat p\hat q=i.
\eeq
Then, interpreting the variables $q,q',q''$ in the STR (\ref{str1}) as
the eigenvalues of the operator $\hat q$, one can write the STR in the form:
\beq\label{lybe1}
w(-1/r'_3|\hat p)w(r'_2|\hat q)w(-1/r'_1|\hat p)=
w(r_1|\hat q)w(-1/r_2|\hat p)w(r_3|\hat q),
\eeq
where the function $w(r|q)$ is defined in (\ref{wrq}). A particular choice
of the parameters, $r_i'r_i=-1$, $r_1=r_3=-r_2=\pm1$, in (\ref{lybe1})
gives a constant solution to the YBE, which is discussed in \cite{F1}
in the context of the quantum discrete sine-Gordon model.

To interpret (\ref{lybe1}) as a  local YBE, introduce
 a linearly independent set of vectors in $R^3$, $\{e_1,e_2,e_3\}$,
and the 3-dimensional lattice, generated by these vectors with integer
coefficients:
\beq\label{lattice}
M=\{n=n_1e_1+n_2e_2+n_3e_3\ |\ n_1,n_2,n_3\in Z\}
\eeq
Let $a_1,a_2,a_3$ be the functions on $M$:
\[
a_j\colon M\ni n\mapsto a_j(n)\in C,\quad j=1,2,3.
\]
The form of the identity (\ref{lybe1}) suggests the following parametrization:
\[
r'_1=-1/a_1(n),\ r'_2=-a_2(n-e_1-e_3),\ r'_3=-1/a_3(n),
\]
\[
r_1=a_1(n-e_2-e_3),\ r_2=1/a_2(n),\ r_3=a_3(n-e_1-e_2).
\]
Substituting these expressions into (\ref{lybe1}), we obtain the local YBE
as is defined in \cite{MN1,MN2}:
\begin{eqnarray}\label{lybe2}
w(a_3(n)|\hat p)w(-a_2(n-e_1-e_3)|\hat q)w(a_1(n)|\hat p)\nonumber\\
=
w(a_1(n-e_2-e_3)|\hat q)w(-a_2(n)|\hat p)w(a_3(n-e_1-e_2)|\hat q).
\end{eqnarray}
The star-triangle transformation formulae (\ref{stt}) acquire now the
form of difference equations:
\beq\label{discsys}
\frac{a_j(n-e_k-e_l)}{a_j(n)}=
-\frac{1+a_2(n-e_1-e_3)(a_1(n)+a_3(n))}{a_1(n)a_3(n)},
\ \{j,k,l\}=\{1,2,3\}.
\eeq
\begin{theo}
The substitution
\[
a_j(n)=\alpha_j^{-1}
\frac{\tau(n+e_k)\tau(n+e_l)}{\tau(n+e_1+e_2+e_3)\tau(n-e_j)},\quad
\{j,k,l\}=\{1,2,3\},
\]
where $\alpha_j$ are some numerical coefficients,
reduces system of equations (\ref{discsys}) to the only equation on the
function
$\tau(n)$:
\beq\label{Miwa}
\sum_{j=1}^4\alpha_j\tau(n+e_j)\tau(n-e_j)=0,\quad e_4=-e_1-e_2-e_3,
\quad \alpha_4=\alpha_1\alpha_2\alpha_3.
\eeq
\end{theo}
The proof is through the straightforward substitution. Miwa in \cite{M}
has shown that equation (\ref{Miwa}) in the suitable continuous limit
reduces to the BKP equation, while it itself is equivalent to the whole
BKP hierarchy.
The limit $\epsilon\to0$, where $\alpha_j=\beta_j\epsilon$, $\beta_j$
being fixed, reduces equation (\ref{Miwa}) to the Hirota's equation of
the discrete
Toda system \cite{H}:
\beq\label{Hirota}
\sum_{j=1}^3\beta_j\tau(n+e_j)\tau(n-e_j)=0.
\eeq

\section{A difference equation associated with the STR in the Ising model}
\label{secising}

In this section we make a similar analysis to that of sect.~\ref{sec1} of
another STR, associated with
the Ising model. The corresponding dynamical system has been introduced
and studied by Korepanov in \cite{K1,K2}, where he has proved that this
system has enough set of conserved quantities to be integrable, and
described how to construct solutions of the finite gap type.

Define the Bolzmann weights of the Ising model as a function $W$:
\beq
W\colon C\times\{1,-1\}\ni (x,\sigma)\mapsto
W(x|\sigma)=1+x+\sigma-x\sigma\in C,
\eeq
The STR has the form of the following identity:
\begin{eqnarray}\label{str3}
\sum_{\sigma''=\pm1}W(x'_3|\sigma\sigma'')
W(x'_2|\sigma'')W(x'_1|\sigma''\sigma')\nonumber\\
=(1+x'_1x'_2x'_3)W(x_1|\sigma)W(x_2|\sigma\sigma')W(x_3|\sigma'),
\end{eqnarray}
where
\beq\label{isingstt}
\quad x_jx_k=(x'_jx'_k+x'_l)/(1+x'_1x'_2x'_3),\quad \{j,k,l\}=\{1,2,3\}.
\eeq
To rewrite STR (\ref{str3}) in a matrix form, first define the algebra
$C[Z,X]$ of 2-by-2 matrices, generated by matrices $Z$ and $X$, satisfying
the relations:
\[
Z^2=X^2=1,\quad ZX=-XZ.
\]
With these definitions, the STR (\ref{str3}) can be rewritten in the form
(see also \cite{MS}):
\begin{eqnarray}\label{ybe3}
W((1-x'_3)/(1+x'_3)|X)W(x'_2|Z)W((1-x'_1)/(1+x'_1)|X)\nonumber\\
=RW(x_1|Z)W((1-x_2)/(1+x_2)|X)W(x_3|Z),
\end{eqnarray}
where
\[
R=\frac{(1+x_2)(1+x'_1x'_2x'_3)}{(1+x'_1)(1+x'_3)}.
\]
Next, introduce the new parameters through the formulae:
\beq\label{sin}
4\xi_j=(x_j-1/x_j)^2,\quad 4/\xi'_j=(x'_j-1/x'_j)^2,\quad j=1,2,3.
\eeq
Consider three
complex valued functions $ b_j(n)$, $j=1,2,3$, on the 3d lattice $M$
defined in (\ref{lattice}), and identify their values with the above
defined parameters as follows:
\[
\xi_1=1/b_1(n),\quad \xi_2=b_2(n-e_1-e_3),\quad \xi_3=1/b_3(n),
\]
\beq\label{param}
\xi'_1=1/b_1(n-e_2-e_3),\quad \xi'_2=b_2(n),\quad \xi'_3=1/b_3(n-e_1-e_2),
\eeq
Operator identity (\ref{ybe3}) now takes the form of the local YBE.
\begin{theo}
Under the substitution
\[
b_j(n)=-\frac{f(n+e_k)f(n+e_l)}{f(n+e_1+e_2+e_3)f(n-e_j)},\quad
\{j,k,l\}=\{1,2,3\},
\]
the star-triangle transformation formulae (\ref{isingstt}) together with
(\ref{sin}) and (\ref{param}) imply the following equation on the function
$f(n)$:
\begin{eqnarray}\label{Kor}
2\sum_{j=1}^4(f(n+e_j)f(n-e_j))^2+
4\prod_{j=1}^4f(n+e_j)+4\prod_{j=1}^4f(n-e_j)\nonumber\\
=\left(\sum_{j=1}^4f(n+e_j)f(n-e_j)\right)^2,\quad e_4=-e_1-e_2-e_3.
\end{eqnarray}
\end{theo}
To prove the theorem one has to exclude the dependent variables by
calculating the corresponding resultants, which in intermediate steps
are cumbersome.
In these calculations the Maple computer system has been used.

Consider now a particular continuous limit of the equation (\ref{Kor}).
Let $\epsilon$
be the lattice spacing. Introduce the field
\beq\label{phi}
\phi(\epsilon n)=3\log(f(n)),\quad n\in M.
\eeq
\begin{theo}
In the limit $\epsilon\to 0$ with $x=(x_1,x_2,x_3)=\epsilon n\in R^3$ fixed,
the equation (\ref{Kor}) for the function (\ref{phi}) is reduced to the
following differential equation:
\beq\label{diff}
6\Delta_2\phi\sum_{j=1}^4(\partial_j^2\phi)^2=
2(\Delta_3\phi)^2+(\Delta_2\phi)^3+8\sum_{j=1}^4(\partial_j^2\phi)^3,
\eeq
where $\phi=\phi(x)$, $x=(x_1,x_2,x_3)$,
\[
\Delta_a=\sum_{j=1}^4\partial_j^a,\quad a=2,3,
\]
\[
\partial_j=\partial/\partial x_j, \quad j=1,2,3;\quad
\partial_4=-\partial_1-\partial_2-\partial_3.
\]
\end{theo}
To prove the theorem one has to expand both sides of (\ref{Kor}) in power
series of $\epsilon$ up to the order of $\epsilon^6$.

Thus, we have obtained a differential equation which should be integrable
as a continuous limit of the discrete integrable equation, and which is
symmetric with respect to any permutation of the three coordinates in $R^3$.

\section{Solutions to the tetrahedron equation}\label{secte}

The quantum theory of discrete systems, associated with the local YBE,
should be
described somehow by solutions to the tetrahedron equation (TE). Here we
show how
one can construct certain operator solutions to the TE using the
star-triangle transformations considered in sections~\ref{sec1} and
\ref{secising}.

First, consider  the case associated with electric networks. Let $R_+$ be the
set of positive real numbers:
\[
R_+=\{r\in R\ |\ r>0\},
\]
and $\iota$ be the standard involution in it:
\beq\label{map}
\iota\colon R_+\rightarrow R_+,\quad \iota^*(r)=1/r,
\eeq
where the variable $r$ is interpreted as a coordinate function on $R_+$.
Define the following rational involution in $R_+^3$:
\beq\label{map1}
\varphi\colon R_+^3\rightarrow R_+^3,\quad
\varphi^*(r_j)=r_j\frac{r_1+r_2+r_3}{r_1r_2r_3},\quad j=1,2,3,
\eeq
where $r_1,r_2,r_3$ are the coordinate functions of the corresponding
multiples
in $R_+^3$. Note, that $\varphi$ commutes with the action of the permutation
group on $R_+^3$. Formulas (\ref{map1}) are derived in fact from the
star-triangle transformation (\ref{stt}) through the following identification:
\[
r'_j=1/\varphi^*(r_j),\quad j=1,2,3.
\]
Define now one more involution of $R_+^3$ as the following composition:
\beq\label{map2}
\Phi=\iota_2\circ\varphi\circ\iota_2\colon R_+^3\rightarrow R_+^3,
\eeq
where $\iota_2=\mbox{id}\times\iota\times \mbox{id}$.
Let $\Phi_{ijk}$, $1\le i<j<k\le 6$, be the involutions of $R_+^6$,
which act as $\Phi$ on the subspace $R_+^3\subset R_+^6$, specified by
the $i$-th, $j$-th
and $k$-th $R_+$-multipliers, and identically, on the others.
\begin{theo}
The following TE holds in $R_+^6$:
\beq\label{te1}
\Phi_{123}\circ\Phi_{145}\circ\Phi_{246}\circ\Phi_{356}=
\Phi_{356}\circ\Phi_{246}\circ\Phi_{145}\circ\Phi_{123}.
\eeq
\end{theo}
The proof is the straightforward verification. The identity (\ref{te1})
implies the operator identity
\beq\label{te2}
\Phi^*_{123}\Phi^*_{145}\Phi^*_{246}\Phi^*_{356}=
\Phi^*_{356}\Phi^*_{246}\Phi^*_{145}\Phi^*_{123}
\eeq
in the linear space ${\cal F}(R_+^6)$ of complex-valued functions on $R_+^6$.
Thus, we have obtained an operator solution to the TE. Solutions of such kind,
called solutions to the ``functional'' TE, were considered also by Korepanov
\cite{K3}.

Let us comment on the limit, corresponding to Hirota's equation (\ref{Hirota}).
For $\lambda\in R_+$ define the ``rescaling'' mapping:
\[
g_\lambda\colon R_+^3\rightarrow R_+^3,\quad g^*_\lambda(r_j)=\lambda r_j,
\quad j=1,2,3,
\]
and consider the following composition:
\[
\Phi_\lambda=g_{1/\lambda}\circ\Phi\circ g_\lambda\colon R_+^3\rightarrow
R_+^3,
\]
where $\Phi$ is defined in (\ref{map2}), (\ref{map1}), (\ref{map}). Clearly,
$\Phi_\lambda$ satisfies the TE (\ref{te1}) for any $\lambda\ne0$.
The limit $\lambda\to0$, however,
is non-singular: $\lim_{\lambda\to0}\Phi_\lambda=\Phi_0$, and the
mapping $\Phi_0$, associated with the Hirota's equation (\ref{Hirota}),
satisfies  (\ref{te1}) as well.

Consider now the star-triangle transformation (\ref{isingstt}),
corresponding to the Ising model. Let $I$ be the open unit interval in $R$:
\[
I=\{x\in R\ |\ 0<x<1\},
\]
and $\jmath$ be the following involution in $I$:
\beq\label{smap}
\jmath\colon I\rightarrow I,\quad \jmath^*(x)=(1-x)/(1+x),
\eeq
where $x$ is considered as a coordinate function on $I$.
Define the following involution in $I^3$:
\[
\psi\colon I^3\rightarrow I^3,
\]
\beq\label{smap1}
\psi^*(x_j)=\sqrt{
\frac{(1-x_kx_l)^2-x_j^2(x_k-x_l)^2}{(1+x_kx_l)^2-x_j^2(x_k+x_l)^2}},
\quad \{j,k,l\}=\{1,2,3\}.
\eeq
This mapping is related with (\ref{isingstt}) through the following
identification:
\[
\psi^*(x_j)=(1-x_j')/(1+x_j').
\]
The mapping
\beq\label{map3}
\Psi=\jmath_2\circ\psi\circ\jmath_2\colon I^3\rightarrow I^3,
\eeq
should satisfy the TE identity (\ref{te1}) with $\Phi$'s replaced by $\Psi$'s
and considered in $I^6$. We have verified this identity numerically,
and claim it as a conjecture. The corresponding mapping
\[
\Psi^*\colon {\cal F}(I^3)\rightarrow{\cal F}(I^3)
\]
can be interpreted as a linear operator in the functional space ${\cal
F}(I^3)$.

\section{Summary}

Integrable discrete equations in three dimensions can be obtained
through solutions to the local YBE. Particularly, the star-triangle
transformation in electric networks (\ref{stt}) can be associated with
a solution to the local YBE
(\ref{lybe2}),
which leads to the integrable bilinear Miwa's equation (\ref{Miwa}).
For the Korepanov's dynamical model, related with the star-triangle
transformation in the Ising model, in this way we obtain another
discrete equation (\ref{Kor}). The continuous limit of the latter is
described by a three dimensional differential equation (\ref{diff}),
which is symmetric with respect to any permutation of coordinate axes.

The operator solutions (\ref{map2}), (\ref{map1}),
(\ref{map}), and (\ref{map3}), (\ref{smap1}),
(\ref{smap}) to the TE equation (\ref{te1}), connected with these dynamical
models, may help to quantize the latter. Note, that the quantum version of
Hirota's equation (\ref{Hirota}) has been constructed in \cite{KR} in the
context of the quantum inverse scattering method.

\section{Acknowlegements}

The author is grateful to L.D. Faddeev and I.G. Korepanov for the
stimulating conversations, and to J.M. Maillet for the discussion
of the local YBE. The work is supported by the Programme TEMPRA-Europe
de l'Est from the R\' egion Rh\^ one-Alpes.

\end{document}